\input{aipcheck}

\documentclass[
    ,final            
  ]
  {aipproc}

\layoutstyle{6x9}

\begin{document}

\title{Double Helicity Asymmetry and Cross Section for $\eta$
Production in Polarized pp Collisions at PHENIX}


\author{Frank Ellinghaus\footnote{Frank.Ellinghaus@desy.de}\\ (for the PHENIX Collaboration)}{
  address={University of Colorado, Boulder,
  Colorado 80309-0390, USA}
}

\begin{abstract}
Measurements of double helicity asymmetries for inclusive
hadron production in polarized proton-proton collisions are
sensitive to spin--dependent parton distribution functions, in particular 
to the gluon distribution, $\Delta g$.
This study presents the double helicity asymmetry and the cross section for
$\eta$ production ($\vec{p}+\vec{p} \rightarrow \eta+X$). The cross
section measurement yields valuable data for the extraction of the
fragmentation functions, which are unknown at present but are a
prerequisite for the extraction of the spin--dependent parton distribution
functions.

\end{abstract}

\maketitle


\section{Introduction}
Present knowledge about spin--dependent parton distribution functions (PDFs)
in the nucleon mainly comes from next-to-leading order (NLO) QCD fits
(see, e.g., \cite{bb,deflorian}) to the spin--dependent structure function $g_1$ 
as measured in polarized inclusive deep--inelastic scattering (DIS) experiments (see, e.g., \cite{hermes_g1,compass_g1_d}).
The resulting spin--dependent gluon PDFs have rather large uncertainties due to
the fact that the exchanged virtual photon does not couple directly, i.e., at
leading order, to the gluon. Thus, additional data from polarized pp
scattering in which longitudinally polarized gluons are directly probed via scattering off
longitudinally polarized gluons or quarks should greatly reduce the uncertainties in the NLO
fits, especially for the gluon distribution.

First results from the PHENIX and the STAR collaborations at the Relativistic Heavy Ion Collider
(RHIC) are available,
measuring double helicity asymmetries in inclusive $\pi^0$ \cite {allpi0} and jet 
production \cite{star_jets}, respectively. 
The double helicity asymmetry is defined as 
\begin{equation} \nonumber
A_{LL} = \frac{\sigma^{++}-\sigma^{+-}}{\sigma^{++}+\sigma^{+-}} 
= \frac{\Delta \sigma}{\sigma}, \quad with \quad 
\Delta \sigma
\propto \sum_{abc}\Delta f_a\otimes \Delta f_b\otimes\Delta\hat{\sigma}^{ab\rightarrow cX'}\otimes D^{h}_c
\label{cross_sec_asy}
\end{equation}
where the cross section $\sigma^{++}$ ($\sigma^{+-}$)
describes the reaction where both protons have the same (opposite) helicity. 
The spin--dependent term is given on the right hand side,
where $\Delta f_a$, $\Delta f_b$ represent the spin--dependent PDFs for quarks (u,d,s) and
gluons, and $\Delta \hat{\sigma}$ are the spin--dependent hard scattering cross
sections calculable in perturbative QCD.
The fragmentation functions (FFs)
$D^h_c$ represent the probability for a certain parton $c$ to fragment into a certain
hadron $h$, and thus they are not needed in the case of jet production.
It is apparent that the extraction of the spin--dependent PDFs from, e.g., the
PHENIX $\pi^0$ data, critically depends on the knowledge of the FFs.
It therefore is desirable to extract the spin--dependent PDFs using different channels, not only because of
possibly different systematic uncertainties involved in the $A_{LL}$ measurement, but 
also because of the different FFs involved. The latter can be extracted based on
cross section measurements from independent data, mainly from $e^+ e^-$ colliders.

This study focuses on $A_{LL}$ in $\eta$ production
($\vec{p}+\vec{p} \rightarrow \eta+X$). In contrast to the $\pi^0$ the $\eta$ FFs are
unknown, but at least the quark FFs can be reasonably well constrained based on
the existing $\eta$ cross section measurements from $e^+ e^-$ collider
data. 
The extraction of gluon FFs requires either $e^+ e^-$
data taken in a wide range of $\sqrt s$ values, or cross section measurements
from, e.g., pp scattering. The $\eta$ cross section has been recently extracted by the
PHENIX collaboration based on the data taken in 2003 \cite{eta_long}. An extraction based on
the data from 2005, with improved statistics and a wider range in transverse
momentum is in progress. It will serve as an important input for the $\eta$ FF extraction which
is in preparation \cite{marco}.

Apart from serving as input in a global fit in order to constrain
the spin--dependent gluon PDF, the $A_{LL}$ in $\eta$ production might have a
second interesting application. Due to the $s$-quark content in
the $\eta$ wave function which is absent in the $\pi^0$, it is expected
that the fragmentation from $s$-quarks into $\eta$ mesons is larger than the one
into $\pi^0$ mesons. If the $s$-quark contribution turns out to be significant,
the $A_{LL}$ in $\eta$ production might help to constrain the spin--dependent
$s$-quark PDF, either in a global fit or by a dedicated analysis 
focusing on possible differences between the $A_{LL}$ in $\pi^0$
and $\eta$ production.

\section{Reconstruction of $\eta$ mesons}
The $\eta$ meson is
reconstructed via its main decay channel $\eta \rightarrow  \gamma \gamma$ with a branching
ratio of about 40\%. The three-body decay $\eta \rightarrow \pi^+ \pi^- \pi^0$
not only has a smaller branching ratio of about 23\%, but in addition also has
a smaller acceptance in the PHENIX spectrometer and therefore has not been
considered yet.
The data were taken at the PHENIX \cite{phenix} experiment in 2005. The 
primary detector used in this analysis is the electromagnetic calorimeter (EmCal),
located at a radial distance of about 5~m from the beam pipe.
It covers the pseudo--rapidity range $|\eta| < 0.35$ and has an azimuthal
acceptance of $\Delta \phi = \pi$.
The EmCal consists of eight sectors, six of which are composed of a total of 15552
lead--scintillator (PbSc) sandwich modules 
(5.5~cm x 5.5~cm x 37.5~cm), 
and two sectors of lead-glass (PbGl) cherenkov calorimeters, consisting of a
total of 9216 modules (4~cm x 4~cm x 40~cm).

A cluster in the EmCal is assumed to originate from a photon if the following
criteria are met. First, the cluster may not be associated with a signal from a charged particle in
the Pad Chamber just in front of the EmCal (PC3); an exception is made if
the signal position in the EmCal and in the PC3 are aligned in such a way that the particle
likely came from the vertex on a straight line, i.e., it was not bent in the
central magnetic field. In this case the cluster is accepted as a photon
candidate since it is assumed that the original photon converted into an
$e^+ e^-$ pair just before the PC3.
Furthermore, since electromagnetic showers in the EmCal are not confined within a
single module, a shower profile analysis can be used in order to reject hadrons
which usually produce broader showers than photons. Since hadrons are
slower than photons, an additional time of flight cut is used for the photon
identification. 

\begin{figure}[h]
\includegraphics[width=0.85\columnwidth]{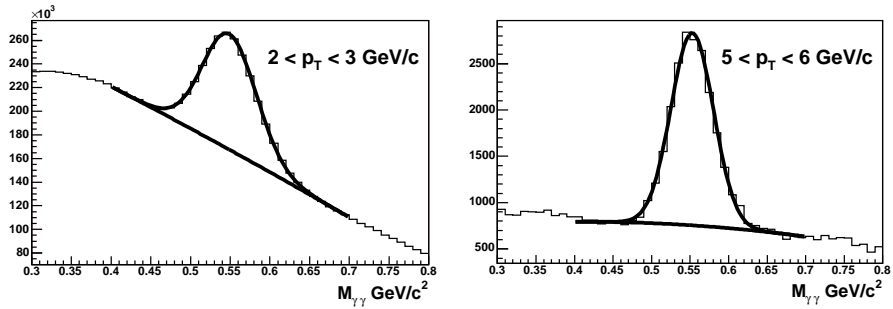}
\caption{Two photon invariant mass distribution for two different $p_T$
  bins.}
\label{invmass}
\end{figure}

Using all possible pairs of photon candidates, the two--photon invariant mass can be
calculated. In order to exclude clusters
with potentially wrongly reconstructed energies from the calculation, 
the module with the largest energy deposition in a cluster may not be in the
outermost two columns or rows of an EmCal sector, and there may not be a noisy
or dead module in the eight surrounding modules. 
An upper limit of 0.7 is placed on the energy asymmetry $(E_1 - E_2)/(E_1 + E_2)$
of the two cluster energies $E_1$ and $E_2$ in order to reduce the
combinatorial background. Finally, the transverse momentum $p_T$ of the di--photon is required
to be larger than 2~GeV/c. 
The invariant mass distribution for two different $p_T$ bins is shown in Fig.\ref{invmass}. In the
vicinity of the $\eta$ peak the distribution is well
described by a fit to a Gaussian plus a second--order polynomial. 
For the asymmetry calculation the events within
a $2\sigma$ window around the mean of the $\eta$ peak are used, whereby the number of
background events $N^{BG}$ is given by the integral under the polynomial within
this window.

\section{Double Helicity Asymmetry}
Experimentally, the double helicity
asymmetry (Eqn.~\ref{cross_sec_asy}) translates into
\begin{equation} \nonumber
A_{LL} = \frac{1}{|P_B||P_Y|}\frac{N_{++}-RN_{+-}}{N_{++}+RN_{+-}}, \quad
with \quad 
R\equiv\frac{L_{++}}{L_{+-}},
\end{equation}
where $N_{++}$ ($N_{+-}$) 
is the experimental yield for the case where the beams
have the same (opposite) helicity.
The relative luminosity $R$ is measured by a coincident signal in two beam--beam counters, which 
have a full azimuthal coverage at a distance of about $\pm 1.4$~m from the target.
The achieved uncertainty on $R$ is on the order of $10^{-4}$.
The polarizations of the two colliding beams at RHIC are denoted by $P_B$ and $P_Y$. 
The degree of polarization is determined from the combined information of a
$\vec {p}C$ polarimeter~\cite{pC}, using an unpolarized ultra--thin
carbon target, 
and from $\vec {p}\vec{p}$ scattering, using a polarized atomic
hydrogen gas-jet target~\cite{jet}. 
The preliminary average polarization value for the data from 2005 is $47$\%
with an uncertainty of $20$\% per beam, leading to a 40\% scale uncertainty
in the preliminary $A_{LL}$ result.

\begin{figure}[h]
\includegraphics[width=0.75\columnwidth]{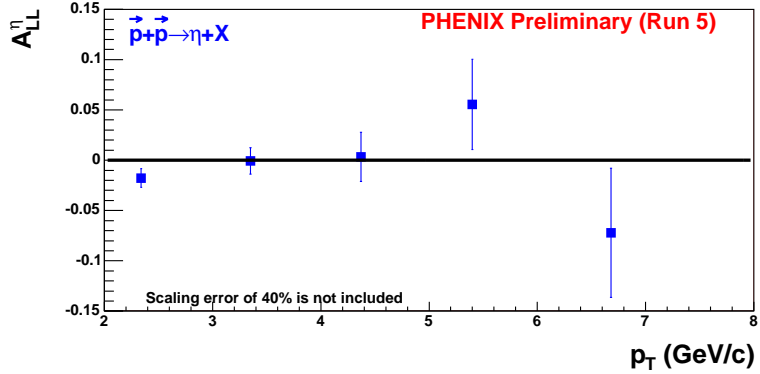}
\caption{Double helicity asymmetry for the inclusive $\eta$ production as a
  function of $p_T$.}
\label{asy}
\end{figure}

The double helicity asymmetry for $\eta$ production is calculated as
\begin{equation} \nonumber
A_{LL}^{\eta} = \frac{A_{LL}^{\eta+BG}-rA_{LL}^{BG}}{1-r}, \quad with  \quad
r\equiv\frac{N^{BG}}{N^{\eta}+N^{BG}},
\end{equation}
where $A_{LL}^{\eta + BG}$ is the asymmetry from the events in the $2
\sigma$ window around the mean of the $\eta$ peak, 
and $A_{LL}^{BG}$ is the background asymmetry calculated in an invariant mass 
region below (300--400 MeV/c$^2$) and above (700--800 MeV/c$^2$) the $\eta$
peak (see Fig.~\ref{invmass}). The
asymmetries in the two regions are consistent with each other and therefore
have been combined to form $A_{LL}^{BG}$. The ratio $r$ gives the number of background events
divided by all events in the $2 \sigma$ window around the mean of the $\eta$
peak. The background corrected asymmetry $A_{LL}^{\eta}$ as a function of
$p_T$ is shown in Fig.~\ref{asy}.
As discussed above, the asymmetry can only be related to the spin--dependent
gluon distribution after the $\eta$ fragmentation functions have been extracted.
Similar to the PHENIX results on the $\pi^0$ asymmetry, the $\eta$ asymmetry is
consistent with zero over the measured range.
It should be noted that the data taken in 2006 will
improve the statistical error for $A_{LL}^{\eta}$ by about a factor of three.   \\

This work is supported in part by the US Department of Energy.



\begin{thebibliography}{9}
\bibitem{bb} J. Bl\"umlein, H. B\"ottcher, Nucl. Phys. B 636 (2002) 225.
\bibitem{deflorian} D. de Florian, G.A. Navarro, and R. Sassot, Phys. Rev. D
  71 (2005) 094018.
\bibitem{hermes_g1} [HERMES Collaboration], A. Airapetian et al., Phys. Rev. D (in press), hep-ex/0609039.
\bibitem{compass_g1_d} [COMPASS Collaboration], V.Yu. Alexakhin et al., hep-ex/0609038.
\bibitem{allpi0} [PHENIX Collaboration], S.S. Adler et al., Phys. Rev. Lett. 93
(2004) 202002. S.S. Adler et al., Phys. Rev. D 73 (2006) 091102. 

\bibitem{star_jets} [STAR Collaboration], B.I. Abelev et al., submitted to
  Phys. Rev. Lett., hep-ex/0608030.

\bibitem{eta_long} [PHENIX Collaboration], S.S. Adler et al., submitted to
  Phys. Rev. C, nucl-ex/0611006.

\bibitem{marco} M. Stratmann, private communication.

\bibitem{phenix} [PHENIX Collaboration], K. Adcox et al., Nucl. Instrum. Meth. A 499 (2003) 469.
\bibitem{pC} O. Jinnouchi et al., RHIC/CAD Note 171 (2004).
\bibitem{jet} H. Okada et al., Phys. Lett. B (in press), hep-ex/0601001.


\end{thebibliography}
\end{document}